\documentclass{jnmp}

\usepackage{amsmath}
\usepackage{graphicx}

\JNMPnumberwithin{equation}{section}

\newtheorem{theorem}{Theorem}

\theoremstyle{definition}
\newtheorem{definition}{Definition}

\begin{document}

\renewcommand{\evenhead}{S~Gladkoff, A~Alaie, Y~Sansonnet and M~Manolessou}
\renewcommand{\oddhead}{The Numerical Study of the Solution
  of the $\Phi_0^4$ Model}

\thispagestyle{empty}

\FirstPageHead{9}{1}{2002}{\pageref{gladkoff-firstpage}--\pageref{gladkoff-lastpage}}{Article}

\copyrightnote{2002}{S~Gladkoff, A~Alaie, Y~Sansonnet and M~Manolessou}

\Name{The Numerical Study of the Solution\\
  of the {\mathversion{bold}$\Phi_0^4$} Model}

\label{gladkoff-firstpage}

\Author{S~GLADKOFF~${}^{\dag^1}{}^{\dag^2}$, A~ALAIE~${}^{\dag^1}{}^{\dag^3}$,
Y~SANSONNET~${}^{\dag^1}{}^{\dag^4}$ and M MANOLESSOU~${}^{\dag^1}$}

\Address{${}^{\dag^1}$~E.I.S.T.I., Avenue du Parc,   95011 
Cergy-Pontoise-Cedex, France\\[10pt]
${}^{\dag^2}$~Cap Gemini Telecom France,  20 Avenue Andr\'e Prothin, 92927
La D\'efense, France\\[10pt]
${}^{\dag^3}$~SITA - 26 Ch.de Joinville, 1216  Geneva, Switzerland \\[10pt]
${}^{\dag^4}$~AAPT - 259 George Street, Sydney  NSW  2000, Australia}

\Date{Received April 03, 2001; Revised July 04, 2001;
Accepted July 05, 2001}

\begin{abstract}
\noindent
We present a numerical study of the nonlinear system
  of $\Phi^4_0 $  equations of motion.
The solution is obtained iteratively,
  starting from a precise point-sequence of the
  appropriate Banach space, for small values of the coupling constant.
The numerical results are in perfect agreement
  with the main theoretical results established
  in a series of previous publications.
\end{abstract}

\section{Introduction}

\subsection{A new non perturbative method}

Several years ago we started a program for
  the construction of a non
  trivial $\Phi^4_4$  model consistent with the
general principles of a Wightman Quantum Field
  Theory
({\it Q.F.T.})~\cite{(Q.F.T.)}.
In references~\cite{MM1} we have introduced
  a non perturbative method for the
  construction of a non trivial solution of
  the system
  of the $\Phi^4$ equations of motion
for the Green's functions, in the Euclidean space
  of zero, one and two dimensions.
  In references~\cite{MM2}
we applied an extension of this method to the case
  of four (and a fortiori of three)-dimensional
Euclidean momentum space.

This method is different in approach from the work
  done in the Constructive {\it Q.F.T.} framework of
  Glimm--Jaffe and others~\cite{G.J.}, and the methods
  of Symanzik who created the basis for a pure
  Euclidean approach to {\it Q.F.T.}~\cite{Sym}.

It is based on the proof of the existence
and uniqueness of the solution of the
corresponding infinite system of dynamical equations
  of motion verified by the sequence of the
Schwinger functions. This solution is obtained
inside a particular subset characterised by alternating
signs and ``splitting'' or factorization properties.

                 The reasons that motivated us for a study in smaller
  dimensions and
  not directly in four, were
  the absence of the difficulties due to the renormalization
  and the pure
  combinatorial character of the problem in zero dimensions.

  Another useful aspect  of the zero dimensional case is the fact that
  it provides a direct way to
test numerically the validity of the method.

         In a more recent work~\cite{GASM} a numerical
  analysis of the $\Phi^4_0$
  solution has been
  obtained following our method.

This paper constitutes a new version of this work.
More precisely we present the numerical study and construction
  of the solution  of the~$\Phi^4$ zero-dimensional problem
realized by
iteration. Our numerical results are coherent with our
  basic theorems. More precisely:
  \begin {enumerate}
\itemsep0mm
\item{} The validity of the contractivity
  criterion by
  the iterated
  mapping ${\mathcal M}^*$, is verified up to the value $\Lambda = 0.01$
for the coupling
  constant. The calculations give an
  indication that it should be true beyond this value,
the limit being below
$0.05$ with the relaxation method used here. An \emph{unstable}
behaviour appears with amplified oscillations for bigger values of
  the coupling constant.

  \item{} The  ``good'' properties
of ``alternating signs''
  and ``splitting'' are certainly verified by the constructed solution
  (truncated sequence $H$).

\item{} When the coupling constant
increases beyond the critical
  value  $\Lambda = 0.05$, the number of
  iterations needed for the convergence $\nu_{\rm conv}$
   grows rapidly and the stability of the Green's functions
  in a finite time is
  ensured only for a reduced number of Green's functions
\end {enumerate}

\subsection{The existence and uniqueness of the
  {\mathversion{bold}$\Phi^4_0$} solution.\\ The theoretical background}

\subsubsection{The vector space {\mathversion{bold}${\mathcal B}$}
and the {\mathversion{bold}$\Phi^4_0$} equations of motion}

\begin{definition}[The space ${\mathcal B}$]
We consider the vector space ${\mathcal B}$
of the sequences $H=\{ H^{n+1}\} _{n =2k+1;  k\in\mathbb N}$
  by the following:

The functions $H^{n+1}$ belong to the space
  $C^{\infty}(\mathbb R^+ )$
  of continuously
differentiable numerical functions of the variable
  $\Lambda \in \mathbb R^+$
  (which physically represents the coupling constant).

  Moreover, there exists a universal (independent
  of n and of $\Lambda$)
  positive constant $K_0$, such that the following uniform
  bounds are verified:
\begin{gather}
\forall  \; n=2k+1,\ k\in \mathbb N \qquad
\left|H^{n+1}(\Lambda)\right|\leq n!  (K_0)^n\qquad \forall\;
\Lambda \in \mathbb R^+. \label{2.27}
\end{gather}
\end{definition}

We suppose that the system of equations under consideration,
  concerns the sequences of Euclidean connected
  and amputated with respect to the
free propagators Green's functions (the Schwinger functions).
and that these sequences  denoted
by $H=\{ H^{n+1}\}_{n =2k+1, k\in\mathbb N}$ belong
to the above space~${\mathcal B}$.

Taking into account the facts that in the present
  zero-dimensional
  case all the external
  four-momenta are set equal to zero,
  that the physical mass
  can be taken equal to $1$
  and that the
  renormalization parameters must be set equal
  to their trivial values
  one directly obtains
  the corresponding
  infinite system of equations of
  motion
  for the sequence of the Schwinger functions
  in the following form:
\begin{equation}
\forall \; \Lambda \in\mathbb R^{+} \qquad
       H^2 (\Lambda) =  - \Lambda H^4(\Lambda)  +1
\end{equation}
  and for all $n \geq3$,
\begin{equation}
H^{n+1}(\Lambda) =  A^{n+1}(\Lambda) +
   B^{n+1}(\Lambda) +  C^{n+1}(\Lambda)
  \end{equation}
with:
\begin{gather}
A^{n+1}(\Lambda) =  - \Lambda H^{n+3}(\Lambda);\\
B^{n+1}(\Lambda) =  - 3\Lambda\sum_{\varpi_n(J)}
\frac{n!}{j_{1}!j_{2}!} H^{j_{2}+2}(\Lambda)
H^{j_{1}+1}(\Lambda);\\
C^{n+1}(\Lambda) = - 6\Lambda\sum_{\varpi_n(I)}
\frac{n!}{i_{1}!i_{2}!i_{3}! \sigma_{\rm sym}(I)}
\prod_{l=1,2,3}H^{i_{l}+1}
(\Lambda).
\end{gather}

\subsubsection{The subset {\mathversion{bold}$\Phi \subset{\mathcal B}$}
  and the new contractive mapping {\mathversion{bold}${\mathcal M^ *}$}}

\begin{definition} We introduce the class ${\mathcal D}$ of sequences
\[
\delta= \{\delta_{n} \}
_{n=2k+1; k\in\mathbb N}\in {\mathcal B},
\]
  such that they verify the bounds (\ref{2.27})
  in the following simpler
  form:
\begin{equation}
|\delta_{n} (\Lambda)|\leq  K_0 \qquad \forall\;
  \Lambda \in \mathbb R^+ ,\qquad\forall\;  n=2k+1;  \ k\in \mathbb N.
\end{equation}
\end{definition}

\begin{definition}[Signs and splitting]
A sequence $H \in {\mathcal B}$
belongs to the subset $\Phi \subset{\mathcal B}$
  if there exists a slowly increasing associated sequence
  of positive and bounded functions on~$\mathbb R^+$,
\[
\delta= \{\delta_{n} \}
_{n=2k+1; k\in\mathbb N}\in {\mathcal D},
\]
  such that  the following
  ``splitting'' (or factorization) and sign
  properties are verified $\forall \; \Lambda \in \mathbb R^+$:
\begin{align}
\Phi.1\ \ &   H^2 (\Lambda) =  1  +\Lambda \delta_1
(\Lambda),\quad \mbox{with:}\  \ \lim_{\Lambda \rightarrow 0}
\delta_1 (\Lambda) = 0;\\ \Phi.2 \ \ &     H^4 (\Lambda) =  -
\delta_3 (\Lambda) \left[H^2 (\Lambda)\right]^3,\quad
\mbox{with:}\  \
   \delta_3 (\Lambda)\leq 6\Lambda ,\  \  \lim_{\Lambda \rightarrow 0}
\frac{\delta_3 (\Lambda)}{\Lambda} = 6;\\
\Phi.3 \ \ &
\forall \; n\geq 5 \qquad
         H^{n+1} (\Lambda) = \frac{\delta_n (\Lambda)
         C^{n+1}}{3\Lambda n(n-1)},  \quad  \mbox{with:}\  \
\lim_{\Lambda \rightarrow 0} \frac{\delta_n (\Lambda)}{\Lambda} =  3n(n-1).
\end{align}
  Moreover $\exists$ a uniform bound at infinity
$\delta_{\infty}^{\Lambda}$
  such that:
\begin{equation}
\lim_{n \rightarrow \infty}\delta_{n}(\Lambda)\leq \delta_{\infty}^{\Lambda}.
\end{equation}
\end{definition}

\begin{definition}[The new contractive mapping ${\mathcal M^*}$]
We define the following application
  ${\mathcal M^*}: \Phi \stackrel{\mathcal M^*}\longrightarrow
  {\mathcal B}$ by:
\begin{gather}
H^{2'} (\Lambda) = 1+\Lambda \delta_1^{'}(\Lambda)\qquad
\mbox{with}\qquad \delta_1^{'}(\Lambda)= -H^4 (\Lambda),
\\
H^{4'} (\Lambda) = - \delta_3^{'}(\Lambda)\left[H^{2'}\right]^3\quad
\mbox{with}\quad
\delta_3^{'}(\Lambda)= 6\Lambda\! \left[1+6\Lambda H^2\!\left(\frac 32
- 
\frac{\left|H^6\right|}{6\left|H^4\right|\left|H^2\right|}\right)\!\right]^{-1}\!\!\!\!\!\!\!
\end{gather}
and for every $n\geq 5$
         \begin{equation}
H^{n+1'}(\Lambda) = \frac{\delta_n^{'}(\Lambda)
  C^{n+1'}( \Lambda)}{3\Lambda n (n-1)};\end{equation}
         with:
\begin{equation}\delta_n^{'}(\Lambda)=\frac{3\Lambda n(n-1)}
  {1+D_n(H)}
\end{equation}
  and
  \begin{equation}
D_n(H)=\frac{\left|B^{n+1}\right| - 
\left|A^{n+1}\right|}{\left|H^{n+1}\right|}.
\end{equation}
\label{Def.1.4}
\end{definition}

\begin{theorem} {\rm (The contractivity of the mapping ${\mathcal 
M^*}$ inside $\Phi$
and the construction of the unique non trivial $\Phi_0^4$ solution.)}
\vspace{-2mm}
\begin{enumerate}
\itemsep0mm
\item[\rm i.] The subset $\Phi$ is a nonempty, closed, complete subspace of
${\mathcal B}$
in the induced topology.
\item[\rm ii.] There exists a finite positive constant 
$\Lambda^*(\approx 0.01)$
  such that the mapping ${\mathcal M^*}$ is contractive inside
  the subset $\Phi$ and therefore the mapping ${\mathcal M }$
  has a unique nontrivial
  fixed point inside the subset $\Phi$ when $\Lambda\in]0, \Lambda^*]$.
\item[\rm iii.] The unique nontrivial solution of the $\Phi_0^4$
  equations of motion lies in a neihbourhood of the fundamental sequence
  $H_0$.
\end{enumerate}
\end{theorem}

  \section{The numerical study of the {\mathversion{bold}$\Phi^4_0$} solution}

\subsection{The algorithm for the numerical construction
  of the  {\mathversion{bold}$\Phi^4_0$} solution}

\subsubsection{The iterative procedure}
For the numerical realization of the solution of the system,
  we applied the following  iterative procedure:
We considered the so called ``fundamental sequence'' $H_0$
mentioned
  in the previous section.
  We have chosen it because:

\vspace{-2mm}

\begin{enumerate}
\itemsep0mm
\item[a.] It has a simple form which ``imitates'' the mapping.

\item[b.] It is a nontrivial point of the appropriate
subset $\Phi$ (validity of the signs and splitting
  properties) where we have shown that the solution is expected to be.
\vspace{-2mm}

\end{enumerate}

For these two reasons we could expect a rapid
  convergence of the iteration to the solution
in the neighbourhood of this sequence $H_0$:
  \begin{gather}
         H^2_0 (\Lambda) = 1+6\Lambda^2,\\
H^4_0 (\Lambda) = - 6\Lambda\left[  H^2_0 (\Lambda)\right]^3,
\end{gather}
and $ \forall\;  n\geq 5 $
\begin{equation}
H^{n+1}_0 (\Lambda) =  \frac{\delta_{n, 0}(\Lambda) C^{n+1}_0 (\Lambda)}
{3\Lambda n (n-1)},
\end{equation}
         with:
\begin{equation}C^{n+1}_0(\Lambda) = - 6\Lambda\sum_{\varpi_n(I)}
  \frac{n!}{i_{1}!i_{2}!i_{3}! \sigma_{\rm sym}(I)}
\prod_{l=1,2,3}H^{i_{l}+1}_0
(\Lambda)
\end{equation}
and
\begin{equation}\delta_{n,0}(\Lambda) =
\frac{3\Lambda n(n-1)}{1+0.5\Lambda n\ (n-1) }.
\end{equation}

Then, taking this sequence as starting point we iterated
  the zero-dimensional analog of the mapping
  ${\mathcal M}^*$, precisely Definition~\ref{Def.1.4}.

\subsubsection{Convergence and stopping criteria}

  Let us first introduce the following definitions:

\vspace{-2mm}

\begin{enumerate}
\itemsep0mm
         \item[a.]
  The ``\emph{dimension}'' of each truncated sequence $H$
  used by the computer
  denoted by $n_{\max}$.

\item[b.]
  The sequence $H_{\rm conv}$ which is characterized by all
  components taken at their convergence
  values.

\item[c.]
  The sequence $\delta_{\rm conv}$ which is
the splitting sequence associated with $H_{\rm conv}$.

\vspace{-2mm}

\end{enumerate}

         For some given value of the coupling constant $\Lambda$ and $n_{\max}$,
  the algorithm calculates by iteration the values of $H$
  (that means all the components $H^{n+1}$, for $1\le n\le n_{\max} $),
and stops when the value
  nearly does
  not vary between a step and the next one,
  using the following criteria (independently
  on each component of $H$):
\begin{itemize}
\vspace{-2mm}
\itemsep0mm
\item{} near $0$, (if  $\left|H^{n+1}_{\rm conv} \right|\le \epsilon_{H} $)
  the convergence is obtained if
\[
\left|H^{n+1} - H^{n+1}_{\rm conv} \right| < \epsilon_{H};
\]

\item{} otherwise, by using the value
$ \epsilon_{H}=10^{-10}$
we apply the relative criterium
\[
\left|\frac{H^{n+1}}{H^{n+1}_{\rm conv}}  - 1\right|  <
\epsilon_{H}.
\]
\end{itemize}

\vspace{-2mm}

         When for one component of $H$ the convergence is obtained,
  the value of $H^{n+1}_{\rm conv}$ is frozen,
  and the order of iteration is memorized as $\nu_{\rm conv}$.
         When the convergence is obtained for all the
  components of $H$,
  the algorithm stops,
the values of $H_{\rm conv}$  and  $\nu_{\rm conv}$ are memorized,
  and the values of $\delta$  are
  memorized as  $\delta_{\rm conv}$.
         To avoid an infinite loop, a maximum number
  of iterations is also given.
  Then, the algorithm also stops, and the same
  values are memorized,
  but they are only the last
  obtained values, not the convergence ones,
  since the convergence was not obtained.

\subsection{Graphical representations and conclusions}

\subsubsection{The {\mathversion{bold}$H$}'s and 
{\mathversion{bold}$\delta$}'s as
functions of {\mathversion{bold}$\nu$}}

         We have chosen to realize the computation for $n_{\max} =55$
  and $10$ values of $\Lambda$ from $0.010$
  to $0.100$ because:

\vspace{-2mm}

\begin{itemize}
\itemsep0mm
         \item{}
   Some phenomena do not appear for small values of $n_{\max}$.
   For example if $n_{\max} \le  29$ roughly,
  the convergence is
  obtained for much bigger values of $\Lambda$
  and the oscillations are not observed.

         \item{} For bigger values of $n_{\max}$ the needed
  time and memory space become prohibitive for our
  machines.
\vspace{-2mm}
\end{itemize}

         We obtained the values of $ H^{n+1}$'s
  and  $\delta_n$'s as functions of $\nu$
(the order of iteration) for some given $n_{\max}$ and
$\Lambda$.
  The $ H^{n+1}$'s themselves have (as expected by the method) alternately
  positive $\left(H^{2},H^{6}, H^{10},\ldots\right)$ and negative 
$\left(H^{4}, H^{8}, H^{6},\ldots\right)$
signs (except in the case of divergence).
We find that $ H^2$  converges
  towards values
  a little greater than $1$, (as it was expected theoretically).

         The curves giving $H^{n+1}$'s are very flat, and what
  happens is most clear when we analyse the curves giving
  the
$\delta_n$'s (``splitting''sequences):

For small values of $\Lambda$,
  the algorithm converges monotonically (Fig.~1),
  but damped oscillations
  appear (starting from the value $0.020$)
  for bigger values of
  $\Lambda$. An approximated pseudo-period
  can be found. It is roughly linearly linked
  to the ``dimension'' of the space, because of
  the perturbations due to the particular form of the mapping.

When $\Lambda$ increases, the oscillations
are less and less damped,
  very slightly damped for $0.050$ (Fig.~2),
  and then amplified, more and more, and remain with a smaller and less
stable pseudo-period, before explosion.
         These  oscillations do not appear
  if $n_{\max} \le  29$ roughly, or $\Lambda$ very big.
  In the latter case the divergence occurs before
  the end of the first pseudo-period.

\begin{figure}[t]
\centerline{\includegraphics[width=0.6\linewidth]{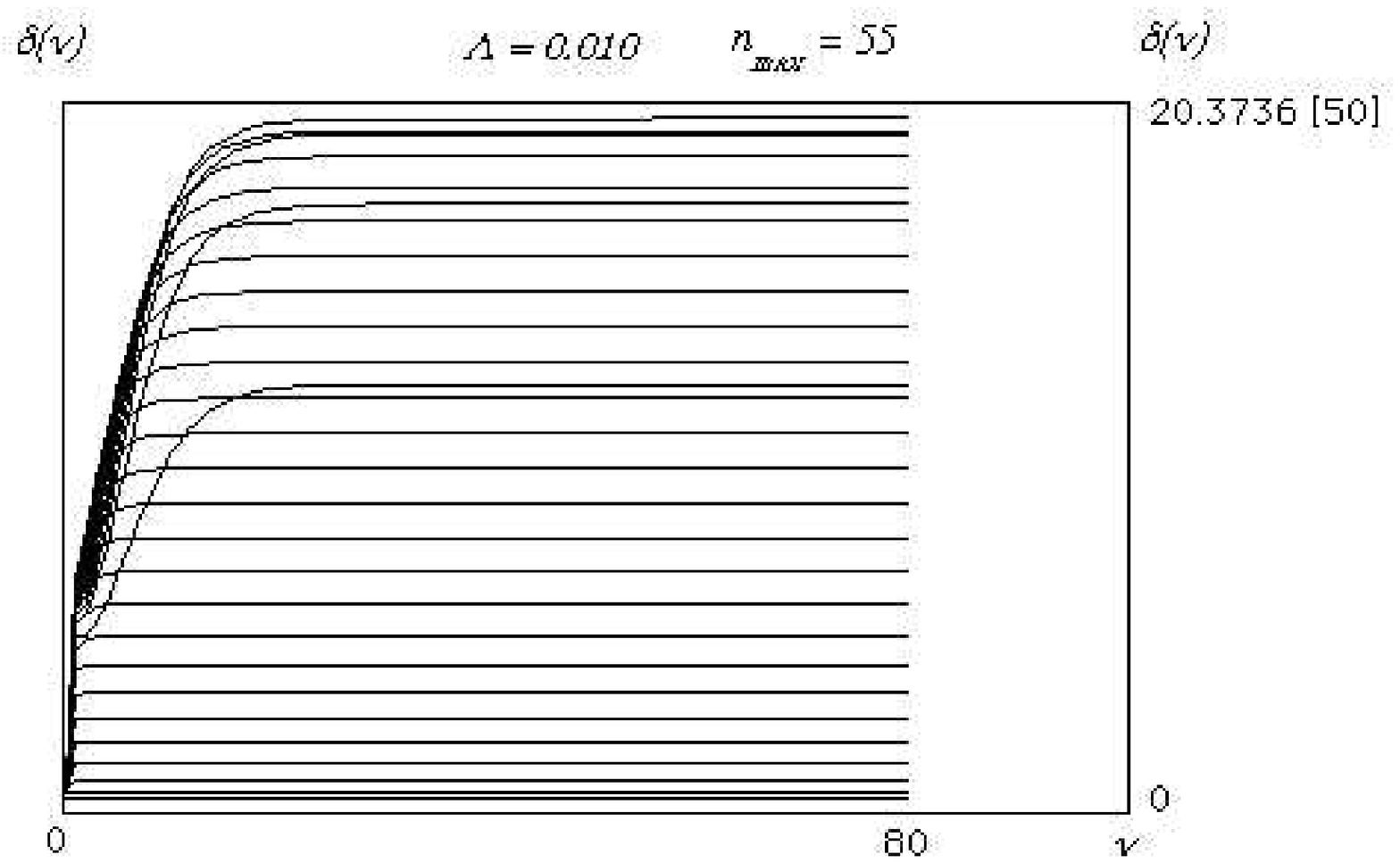}}
\vspace{-2mm}
\caption{Convergence of $\delta_{n;\nu}(\Lambda =0.01)$
as a function of the iteration number $\nu$ for
$ n_{\max} =55$.}

\vspace{4mm}

\centerline{\includegraphics[width=0.6\linewidth]{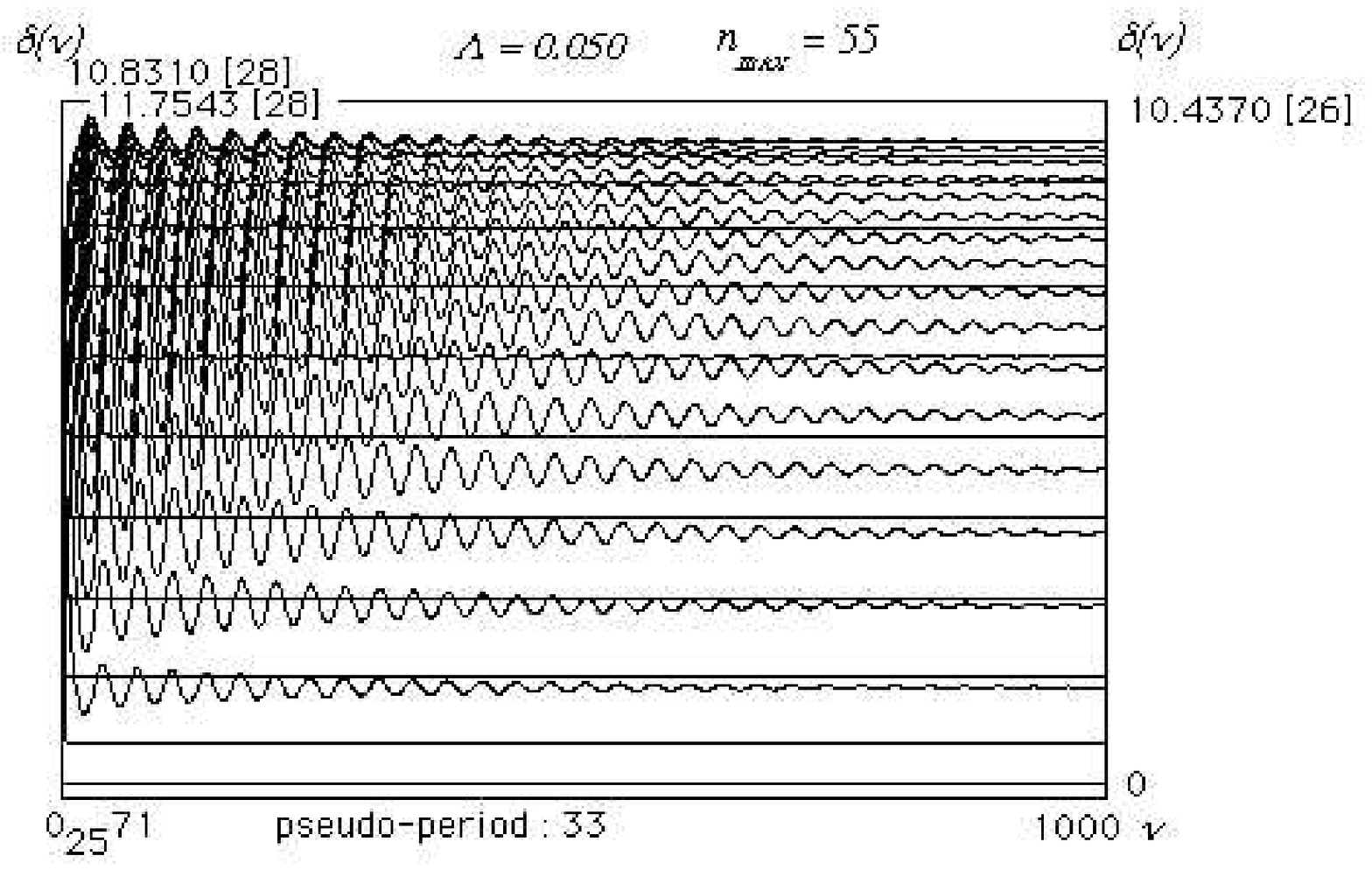}}

\vspace{-2mm}

\caption{Convergence of $\delta_{n;\nu}(\Lambda =0.05)$
as a function of the iteration number $\nu$ for
$ n_{\max} =55$.}
\end{figure}

         The convergence (verification of the contractivity criterium)
which has been proven mathematically up to
  $\Lambda =0.010$ can here be shown numerically
  for  values slightly larger, and we can see how it
is progressively destroyed,
  since a pronounced instability appears
in the form
of amplified oscillations
when the coupling constant increases.

\subsubsection{Curves of {\mathversion{bold}$H_{\rm conv}$},
{\mathversion{bold}$\delta_{\rm  conv}$} and
{\mathversion{bold}$\nu_{\rm conv}$},  as
  functions of {\mathversion{bold}$n_{\max}$}}

In order to keep the article down to
  manageable size we do not present the figures
we comment here. They are, of course,
  at the disposal of all interested readers,
  in the corresponding electronic version of
  the journal~\cite{GASMe}.

\vspace{-2mm}
\begin{enumerate}
\itemsep0mm

\item{} For small coupling constant, $\Lambda\leq 0.03 $,
  the curves of $H_{\rm conv}$
  (i.e.\ the values
  of  $H^{n+1}$'s for which the convergence is obtained)
  as a function of $n_{\max}$
  (the dimension of the used truncated sequence)
show  a good convergence, for sufficiently small $n_{\max}$.
But, when $n_{\max}$
   increases, analogous instabilities
  appear again: there
  are damped oscillations,
  then amplified oscillations and there
is no more convergence. Just before the
  threshold of this phenomenon,
  the number of iterations
  needed for the convergence grows very rapidly.

\item{} The curves of $\delta_{\rm conv}$
(i.e.\ the corresponding $\delta_{n}$'s of
  the iteration for which the convergence
  is obtained) as a function of $ n_{\max}$,
   show similar properties of the
  iteration procedure. For small values
of the coupling constant
the convergence is obtained (for both the
  $\delta_{n}$'s and the enveloppe of the curves)
  as far as $n_{\max}$
increases.

  For $\Lambda = 0.01$,
no instability appears up to
  $ n_{\max} = 119$ (with a limit of $200$ iterations),
  and the maximum $\delta_{\rm conv}$ obtained can be found
  around $30$.
  For $\Lambda = 0.02 $, some instabilities
  appear for $ n_{\max}$
   near $90$, and for $\Lambda \geq 0.06$, they
begin  for  $ n_{\max}\leq 50$.

\item{}
  Finally, the curves representing $\nu_{\rm conv}$ (the iteration
  order for the convergence of each component)   as a
function of $ n_{\max}$ make evident an analogous behaviour.
Moreover, these curves yield some concrete
idea about the speed of convergence
  of the iteration.

For small values of the coupling constant
  ($\Lambda\leq 0.03$),
  the number of iterations needed to obtain
  the convergence grows roughly linearly with
  the dimension of the space $ n_{\max}$ and
  the component number $ n$. The first
  components converge rapidly and independenly of
$ n_{\max}$ because the different Green's functions
  are very weekly coupled and reach their limit
value before the truncation number ($ n_{\max}$) takes
his largest values.

When the coupling constant grows ($\Lambda\geq 0.04 $),
the required number of iterations for the convergence,
$\nu_{\rm conv}$, grows very rapidly and all the components
$H^{n+1}$,
  but the first ones, converge simultaneously being strongly
coupled between them. The convergence is no more
  obtained for bigger values of $ n_{\max}$.

This divergence occurs for smaller and smaller
$ n_{\max}$, when $\Lambda $ increases.

\vspace{-2mm}
\end{enumerate}

\subsection{Outlook and further investigations}
\begin{enumerate}
\vspace{-2mm}

\itemsep0mm
\item [i)] In order to study more precisely the
  behaviour of the convergence, we also obtained
  numerical results and curves showing
  the ratio
\[
\left|\frac{H^{n+1}_\nu}{H^{n+1}_{\nu_{\rm conv}}}\right |
\]
at fixed $\Lambda $ and $n_{\max}$.
The convergence again is evident up to the values
$\Lambda\leq 0.04 $~(\cite{GM}). Oscillations
  appear from $\Lambda\geq 0.05 $ and they are
amplified for larger values of the coupling constant.
\item[ii)] We also explored another aspect
  of the iteration by perfoming a kind
of  ``\emph{map}''~\cite{GM} as follows.

The iteration process starts \emph{upwards}: From the $H^2$
Green's function and $\Lambda$ we calculate the $H^4,
H^6, \ldots, H^{n_{\max}+1} $
Green's functions using the iteration formulas and we verify
  the sign of each of the Green's functions.
If the alternating signs
(positive for $H^2$-negative for $H^4,\ldots$ etc.) are obtained,
  that means the convergence is guaranteed. If the sign
is wrong, the iteration yields values without any physical
  significance; then, the convergence is impossible.

\item[iii)]
In order to accelerate the convergence of the
  iteration we are looking at two types
  of investigations:
\vspace{-2mm}
\begin{itemize}
\itemsep0mm
\item We can ``\emph{redefine}'' each step of
  the iteration by putting together the calculations of $2$
steps of our previous procedure.

\item{} We also can define another starting point of the
  iteration by choosing appropriate and more physical values
  of the splitting sequence, in terms of the values of the
coupling constant. More precisely, if
instead of the fundamental sequence $H_0$, we take as initial
point of the iteration one
  of the previous fixed points (at a given value of $\Lambda$
and $ n_{\max}$) one expects a very rapid convergence
  to the physical solution without chaotic behaviour.
\vspace{-2mm}
\end{itemize}

These two last points are under investigation \cite{GM}.
\vspace{-2mm}
\end{enumerate}

\subsection*{Acknowledgments}
The authors are indebted to O~Lanford for useful discussions and
comments of previous versions of this work.
They are also grateful to B~Grammaticos for a critical reading
of the paper and crucial remarks.
The authors would like also to thank the director of EISTI, N~Fintz
for his encouragements and excellent conditions at EISTI
  during the elaboration of this work.

\label{gladkoff-lastpage}

\end{document}